\documentclass[twocolumn,preprintnumbers,amsmath,amssymb,pra]{revtex4}
\usepackage[colorlinks,linkcolor=blue,anchorcolor=blue,citecolor=blue]{hyperref}
\usepackage{txfonts}
\usepackage{graphicx,hyperref}
\usepackage{dcolumn}
\usepackage{bm}
\usepackage{braket}
\usepackage{upgreek}
\usepackage[mathscr]{eucal}
 \allowdisplaybreaks
\begin{document}
\title{Chaos signatures of current phase transition in a toroidal trap}
\author{Zhiqiang Li$^{1}$}
\author{Xiaoxiao Hu$^{1}$}
\author{Zhao-Yun Zeng$^{2}$}
\author{Yajiang Chen$^{1}$}
\author{Ai-Xi Chen$^{1}$}
\author{Xiaobing Luo$^{1,2}$}
\altaffiliation{Corresponding author: xiaobingluo2013@aliyun.com}
\affiliation{$^{1}$Department of Physics, Zhejiang Sci-Tech University, Hangzhou, 310018, China}
\affiliation{$^{2}$School of Mathematics and Physics, Jinggangshan University, Ji'an 343009, China}

\date{\today}
\begin{abstract}
	In this work we demonstrate how the directed motion of atomic Bose-Einstein condensates in  a toroidal trap can be controlled by applying a  zero-mean oscillatory driving field. We show that due to the self-trapping effect in momentum space, the oscillatory amplitude of the current can be significantly suppressed and a nearly constant directed current can be obtained preserving the initial current values, by decreasing the driving amplitude, even when the atomic interactions are relatively small. We also reveal numerically the mean-field chaos can serve as an indicator of a quantum phase transition between the vanishing current regime and nonvanishing current regime. Our results are corroborated by an effective three-mode model, which provides an excellent account of  the ratchet dynamics of the system. \\

{Keywords:} self-trapping effect, quantum transport, directed current, mean-field chaos
\end{abstract}
\maketitle
\section{Introduction}
The realization of Bose-Einstein condensates (BECs) of dilute atoms at very low temperatures provides an ideal platform for the study of a rich variety of macroscopic quantum phenomena: quantum coherence \cite{M. R. Andrews, Y. Castin, H. Wallis}, macroscopic quantum tunneling \cite{M. R. Andrews, Y. Castin, H. Wallis, H. Lignier, A. Smerzi, M. Ueda}, and quantum superfluidity \cite{M. R. Andrews2, Q. Chen, A. J. Leggett}, among others. Among the many results, the self-trapping effect, discovered in the pioneering work of Ref. \cite{A. Smerzi}, is a seminal one: when it occurs, the Josephson oscillations are blocked and the atoms in the BEC show a highly asymmetric distribution in the potential well due to interatomic interactions \cite{M. Albiez, L. Fu}.  In the large-particle number limit, the dynamics of a BEC is described by a mean-field nonlinear Gross-Pitaeviskii (GP) equation\cite{F. Dalfovo}, which has been successfully applied to the prediction of a wide range of nonlinear dynamics, such as chaotic behaviour \cite{C. Lee, W. Hai, C. Weiss, E.-M. Graefe, M. Hiller,S. Mossmann,Q. Xie, J. Tan, K. Zhang}, nonlinear Landau-Zener tunneling \cite{J. Liu, G.-F. Wang, B. Wu}, solitons and vortices \cite{S. Burger, L. Khaykovich, B. A. Malomed}, quantum droplets \cite{I. Ferrier-Barbut, Z.-H. Luo, G. Semeghini}, etc. Experimentally, with the aid of increasingly sophisticated laser technology and other experimental means, the condensate can be precisely controlled, not only to prepare a particular initial state with high precision, but also to adjust the strength of the nonlinear interaction between the atoms by means of the Feshbach resonance\cite{J. M. Gerton}.

Controllable, fully coherent quantum transport of ultracold atoms is a prerequisite for promising applications of quantum devices. There are several ways to achieve this goal, including what is known as ratchet effect. By a ratchet effect one usually refers to the existence of directed transport in a spatially periodic  system  by adding  time-periodic modulations without net bias. Ratchets have traditionally been found in the study of classical systems \cite{S. Flach, P. Reimann}, which is relevant to the design of nanoscale devices and the understanding of the mechanism underlying biological motors \cite{Y. V. Gulyaev, P. Hanggi, A. Hashemi}.  It is not surprising that  the addition of quantum effects such as tunneling gives rise to new ratchet phenomena and the concept of quantum ratchets has  broadened  this area of investigation \cite{L. Chen, M. V. Costache, A. Kenfack, E. Lundh}. Generally,  in order to achieve directed (ratchet) transport, the associated  space- and time-inversion symmetries have to be broken\cite{S. Denisov, S. Denisov2, C. Grossert}. However, there has been some discussion of the possibility of obtaining long-lasting directed  currents without the need for the breaking of spatio-temporal symmetries\cite{J. Santos, D. Poletti, C. E. Creffield, D. Poletti2}. The ratchet effect can be exploited to achieve controllable coherent quantum transport; for example, the ratchet current can be substantially increased  by tuning specific Floquet states to avoided crossings \cite{S. Denisov2}. Recently, quantum ratchet studies have been extended to open quantum systems \cite{Z. Fedorova, M. Keck, G. Lyu, A. Metelmann, K. Yamamoto, S. Longhi, W.-L. Zhao}. The examples include the dissipation-driven system which generates a persistent current \cite{Z. Fedorova}, and the directed momentum current which exhibits a staircase growth with time in a non-Hermitian kicked rotor model \cite{W.-L. Zhao}.

In combination with a time-periodic perturbation, the nonlinearity resulting from a mean-field treatment of the interactions between the atoms makes it possible to introduce chaotic motion into the BEC system, which leads to an instability of the condensate wave function. Chaotic behaviour is ubiquitous in nonlinear systems and is closely related to tunneling and transport, both in the classical and in the quantum domains. For example, it is known that chaos can be a substitute for disorder in Anderson's scenario and can induce localization for some special systems \cite{D. R. Grempel, M. Santhanam}. In addition, chaos can be harnessed to control the delocalization-localization transition of cold atoms held in the optical lattice \cite{J. Tan, L. Li2008, L. Li, M. Zou}.  A well-known quantum chaos transport mechanism, called chaos-assisted tunneling, can also significantly enhance tunneling and transport between the stable islands separated by the chaotic sea, which is forbidden in the classical phase \cite{M. Arnal, B. Dietz, D. A. Steck}, providing a novel and interesting scheme for transport. More importantly, in the context of the ultracold atomic system, chaos-related phenomena lend themselves to the study of quantum fluctuations beyond the mean-field level, providing a paradigm for studying quantum chaos.

In this paper we consider how a constant (rather than oscillatory) ratchet current can be generated and tuned in a nonlinear system of a Bose-Einstein condensate on a ring driven by a zero-mean time-periodic potential without breaking the spatio-temporal symmetries. We show that, starting from asymmetric initial states in momentum space, the self-trapping effect induced by nonlinearity can block the oscillations between different momentum modes, freezing the asymmetry so as to produce a persistent and stable directed current. In particular, by means of the self-trapping effect in momentum space, we can realize a novel quantum ratchet effect that enables the condensates to flow in the ring at an almost constant velocity by decreasing the driving amplitude or increasing the driving frequency. Specifically, the average current velocity can be set by the initial condition alone, and the high velocity can be expected by the proper preparation of the  initial conditions. We also establish a strong link between the mean-field chaos and the current phase transition, where the average current changes from zero to a non-zero value, by showing numerically that the Lyapunov exponent reaches a maximum at the transition point. The mean-field chaos manifestation of the current phase transition may be used to detect the effects of quantum fluctuations via the dependence of the current transition on the system parameters.
\section{Model}\label{}
We consider the BEC  atoms confined in a toroidal trap with a cross section much smaller than the radius of the trap, $R$, so that the motion is essentially one-dimensional. The condensate is subjected to a time-periodic continuous drive and  its dynamics is well described by the one-dimensional dimensionless Gross-Pitaevskii nonlinear equation ($\hbar=1$),
\begin{align} \label{eq1}
	i\frac{\partial \psi(x,t)}{\partial t}=&\bigg[\frac{\hat{p}}{2}^2+g|\psi(x,t)|^2+\nonumber\\&K\sin(\omega t)\sin(x)\bigg]\psi(x,t),
\end{align}
where $x$ is the azimuthal angle, the momentum operator $\hat{p}=-i\frac{\partial}{\partial x}$, $g = \frac{8NaR}
{r}$ is the scaled strength of the nonlinear
interaction, $N$ is the number of BEC atoms, and $a$ is the s-wave scattering length between the atom-atom elastic collisions. $K$ and $\omega$ denote the strength and the angular frequency of the time-periodic perturbation, respectively.  Due to the periodic boundary conditions $\psi(x)=\psi(x+2\pi)$, the wave function can be expanded in the basis of the momentum eigenstates, $\psi(x,t)=\sum_{n=-\infty}^{\infty} c_n(t)\phi_n(x)$, where $\phi_n(x)=\sqrt{\frac{1}{2\pi}}e^{inx}$ and $n$ is an integer. We choose the initial state as an equally weighted superposition of the momentum eigenstates $\ket{0}$ and $\ket{1}$,
\begin{equation}
	\psi(x,0)=\frac{1}{\sqrt{4\pi}}(1+e^{-ix}).
\end{equation}
Such an initial state can be prepared experimentally using the Bragg pulse, which explicitly breaks time reversal symmetry and carries a non-zero current. The initial state is different from the one in the earlier paper \cite{D. Poletti}, allowing to show how the initial current is maintained and destroyed under the influence of nonlinearity and periodic drive.

To investigate how the system behaves, we evaluate the current flowing in the ring as follows
\begin{equation}
	I(t)=\int_{0}^{2\pi}dx\psi^{*}(x,t)\hat{p}\psi(x,t).
\end{equation}
The asymptotic time-averaged current (TAC), the main physical quantity of interest, is given by
\begin{equation}\label{Orbitequ1}
	\bar{I}=\lim_{t\to \infty}\frac{1}{t}\int_{0}^{t}dt^{\prime}I(t^\prime) .
\end{equation}
We solve the GP equation (\ref{eq1}) numerically using a split operator method. Due to the limited time length of the numerical simulation, we truncated it to 8000 driving cycles, which makes sense for our investigations.

\section{Current phase transition}
\begin{figure}[htp]
	\center
	\includegraphics[width=7cm]{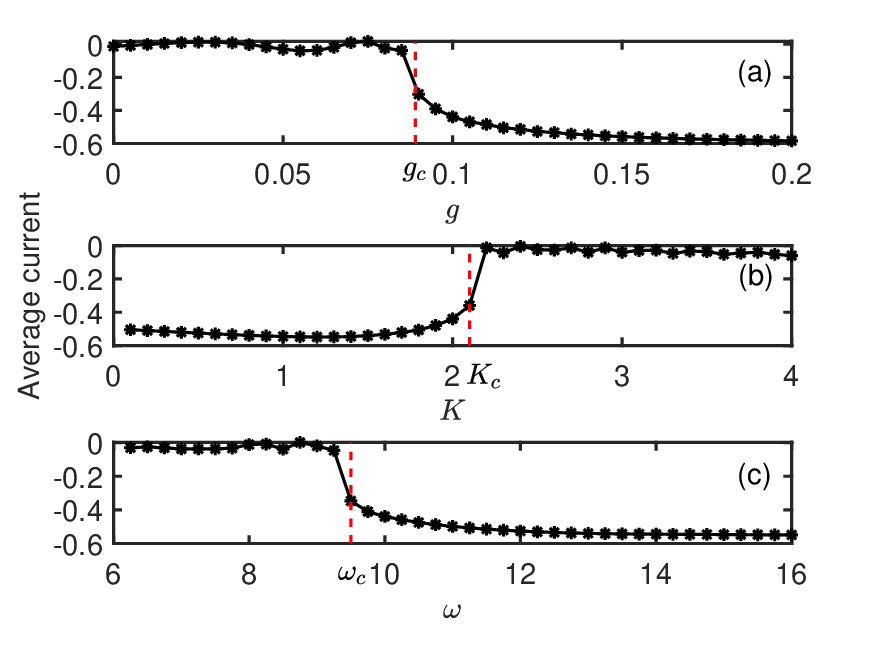}
	\caption{Long-time averaged current versus nonlinearity $g$, driving amplitude $K$ and driving frequency $\omega$. (a): $K=2$, $\omega=10$, the phase transition point $g_c$ is near $g=0.089$. (b): $g=0.1$, $\omega=10$, the phase transition point $K_c$ is close to $K=2.15$. (c): $g=0.1$, $K=2$, the phase transition point $\omega_c$ is near $\omega=9.4$.} \label{q1}
\end{figure}

Through numerical simulations, we find that by varying the nonlinearity strength $g$, or the driving strength $K$, or the driving frequency $\omega$, a phase transition occurs, i.e., the long-time averaged current changes from zero to a non-zero value. As shown in Fig. \ref{q1}(a), the TAC value is practically zero for small nonlinearity strengths, and there is a critical nonlinearity strength ($g_c$) beyond which the average current increases abruptly to a (negative) value almost equal to the initial current. More importantly, we find that such a phase transition, where the average current changes from zero to non-zero, can also be realized by simply decreasing the driving strength $K$ or increasing the driving frequency $\omega$, as illustrated in Fig. \ref{q1}(b) and \ref{q1}(c) .

\begin{figure}[htp]
	\center
	\includegraphics[width=7cm]{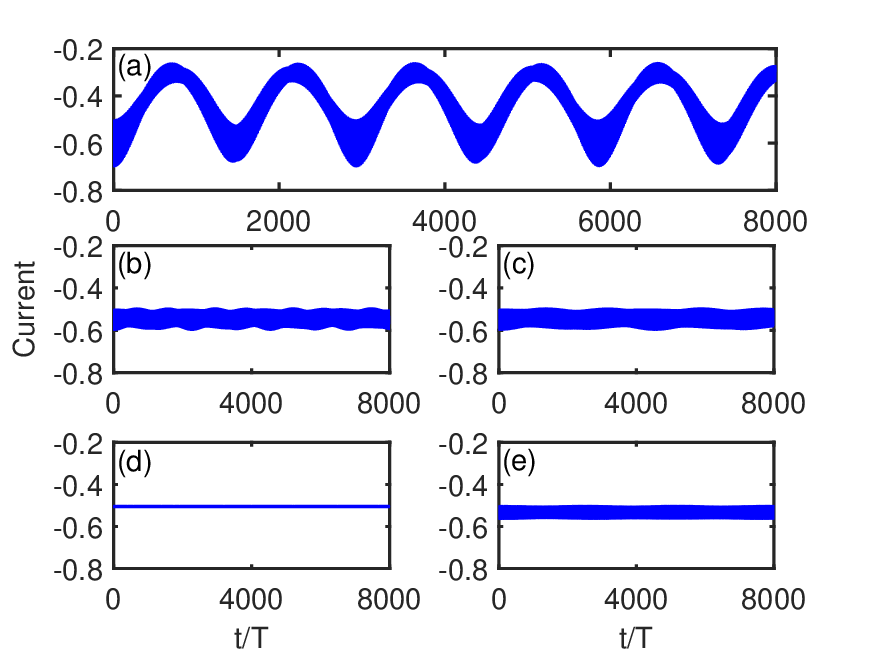}
	\caption{Time evolution of the current. (a): $g=0.1$, $K=2$, $\omega=10$; (b) and (d): the driving amplitude is reduced to $K=1$ and $K=0.1$ respectively, with the other parameters the same as in (a); (c) and (e): the driving amplitude is increased to $\omega=20$ and $\omega=30$ respectively, with the other parameters the same as in (a). } \label{q2}
\end{figure}

Such a phase transition of the current in the nonlinear system, controlled by the driving parameters, offers an alternative option to generate a persistent and stable ratchet current. In Fig. \ref{q2}(a), we set $g = 0.1$ and observe that the current has a large amplitude/long period oscillation in the range of $-0.7\le I(t)< -0.2$, superimposed with a fast, relatively small amplitude oscillation. If we leave $g$ unchanged, and only reduce the driving strength $K$ [see Fig. \ref{q2}(b) and \ref{q2}(d)] or increase the driving frequency $\omega$ [see Fig. \ref{q2}(c) and \ref{q2}(e)], the oscillation range of the current is significantly suppressed, which results in a stable current. Particularly, when $K$ is as low as $0.1$, Fig. \ref{q2}(d) shows a constant value of steady current, implying that condensate flows in the ring at a constant velocity. These results are particularly useful for building a coherent quantum ratchet.

\begin{figure}[htp]
	\includegraphics[width=7cm]{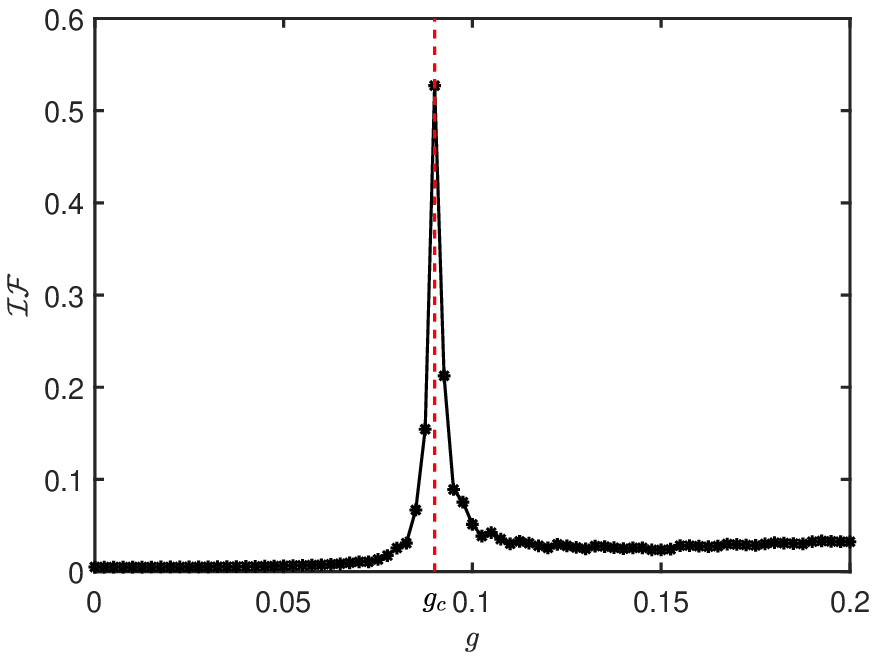}
	\caption{Instability of the mean-field dynamics of the GP equation (\ref{eq1}) with respect to the nonlinearity $g$. The parameters are set to $K=2$, $\omega=10$. At the phase transition point, the $\mathcal{IF}$ defined in Eq. (\ref{eq5}), which is used to quantify the divergence of nearby trajectories, reaches its maximum value.} \label{q3}
\end{figure}

According to our general understanding, the Gross-Pitaevskii equation is nonlinear, and the onset of instability (or chaotic motion) of the condensate could occur, in the sense of exponential sensitivity to the initial condition. It is then curious to check whether the current phase transition can be quantitatively connected with the instability of the mean-field dynamics. To quantify the divergence of nearby trajectories, we introduce
\begin{equation}\label{eq5}
	\mathcal{IF}=1-\min\bigg[\int_0^{2\pi} dx\psi^*(x,t)\tilde{\psi}(x,t)\bigg].
\end{equation}
Here, $\psi(x,t)$ and $\tilde{\psi}(x,t)$ are two time-evolving states starting from $\psi(x,0)$ and its slightly perturbed state $\tilde{\psi}(x,0)$. We trace the evolution of the quantum fidelity between these two states and record its minimum value, which gives us the quantity $\mathcal{IF}$ defined by Eq. (\ref{eq5}). When the BEC is dynamically stable, $\mathcal{IF}$ is close to zero. Conversely, if $\mathcal{IF}$ is greater than zero, the system becomes unstable.  Fig. \ref{q3}
shows how the $\mathcal{IF}$ varies as the nonlinearity strength is increased. We are surprised to see that the $\mathcal{IF}$ exhibits a prominent peak centred exactly on the current phase transition point. The preliminary results indicate a close link between the current phase transition and the instability of the mean-field dynamics.
\section{The mean-field three-mode model}
A detailed examination of the evolution of the system reveals that the only three momentum eigenstates $\ket{0},\ket{\pm1}$ are predominantly excited during the dynamical evolution with the parameters as in Fig. \ref{q1}. To gain a deep insight into the current phase transition, we use the mean-field three-mode ansatz \cite{D. Poletti,C. E. Creffield},
\begin{equation}\label{eq6}
	\psi(x,t)=\frac{1}{\sqrt{2\pi}}[A(t)e^{-ix}+B(t)+C(t)e^{ix}],
\end{equation}
where $A(t)$, $B(t)$ and $C(t)$ are time-dependent, complex coefficients such that $|A|^2+|B|^2+|C|^2=1$. The ansatz (\ref{eq6}) produces the effective mean-field Hamiltonian of the three-mode model (TMM),
\begin{align}\label{eq7}
	H_{\mathrm{eff}}(t)=&\int_{0}^{2\pi}dx\psi^{*}(x,t)\bigg[-\frac{1}{2}\frac{\partial^2}{\partial x^2}+\frac{g}{2}|\psi(x,t)|^2 \nonumber \\
	&+K\sin(\omega t)\sin(x)\bigg]\psi(x,t).
\end{align}
Under the action of the effective Hamiltonian, we obtain the equation of motion for the expansion coefficients $A(t)$, $B(t)$ and $C(t)$,
\begin{align}\label{eq8}
	i\dot{A}(t)=&\frac{g}{2\pi}[(|A(t)|^2+2|B(t)|^2+2|C(t)|^2)A(t)\nonumber\\
	&+B(t)^2C^{*}(t)]+\frac{A(t)}{2}+\frac{iK\sin(\omega t)}{2}B(t),\nonumber\\
	i\dot{B}(t)=&\frac{g}{2\pi}[2(|A(t)|^2+|B(t)|^2+2|C(t)|^2)B(t)\nonumber\\
	&+2A(t)B^{*}(t)C(t)]
	-\frac{iK\sin(\omega t)}{2}[A(t)-C(t)],\nonumber\\
	i\dot{C}(t)=&\frac{g}{2\pi}[(2|A(t)|^2+2|B(t)|^2+|C(t)|^2)C(t)\nonumber\\
	&+A^{*}(t)B(t)^2]+\frac{C(t)}{2}-\frac{iK\sin(\omega t)}{2}B(t).
\end{align}

\begin{figure}[htp]
	\center
	\includegraphics[width=7cm]{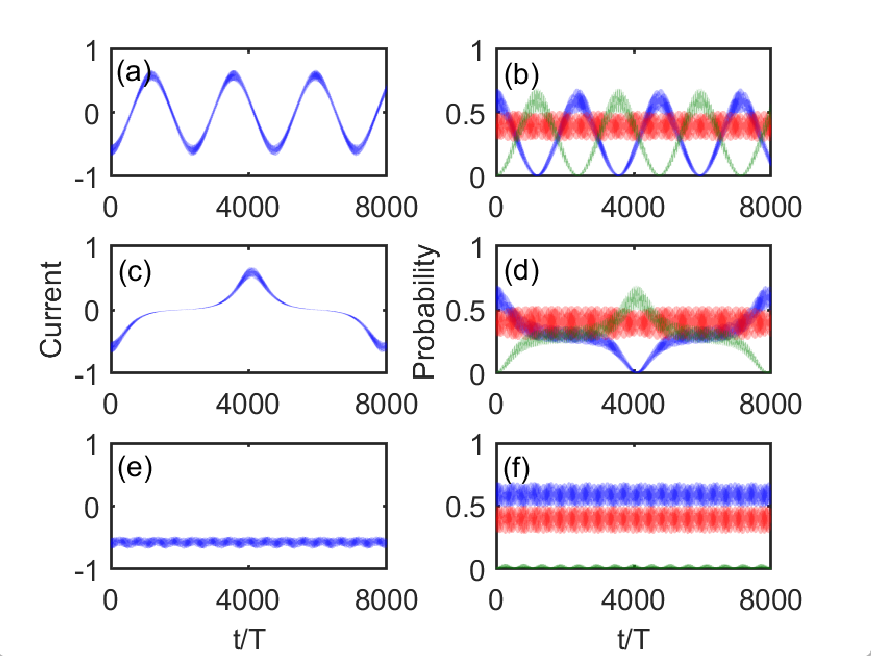}
	\caption{Time evolution of the currents (left column) and the population probabilities  (right column) at the momentum modes $\ket{-1}$ (blue line), $\ket{0}$ (red line) and $\ket{1}$ (green line) predicted by the three-mode model. Top panels: $g=0.05$. Below the phase transition point; middle panels: $g=0.089$, at the transition point. Bottom panels: $g=0.2$, above the transition point. The other parameter values used are $K=2$ and $\omega=10$.} \label{q4}
\end{figure}

This simple TMM does indeed reproduce the current behaviors derived from the simulation of the full GP equation (\ref{eq1}). In Fig. \ref{q4},  with the initial condition $A(0)=B(0)=\frac {1}{\sqrt{2}},C(0)=0$, we show the time evolution of current of the three-mode model for the nonlinearity strength $g$ below, equal to, and above the phase transition point. The three-mode approximation results agree well with the full GP equation simulation shown later (see left column of Fig. \ref{q9}  for GP equation results with the same set of system parameters and initial conditions as in Fig. \ref{q4}). To study the current phase transition more quantitatively, we also plot the time evolution of population probabilities in the three momentum eigenstates, as illustrated in the right column of Fig. \ref{q4}. As the zero momentum state has no contribution to the current, we focus on the oscillation between the two states $\ket{-1}$ and $\ket{1}$. Below the transition point, periodic exchanges occur between $\ket{-1}$ and $\ket{1}$, so that no net current is produced. As the nonlinearity strength approaches the transition point, the period of oscillation between $\ket{-1}$ and $\ket{1}$ gets longer. By increasing the nonlinearity strength beyond the transition point, we find that the oscillations between $\ket{-1}$ and $\ket{1}$ are suppressed and the system becomes self-trapped in the initial momentum distribution. This implies that the current is maintained at a constant level as a result of the self-trapping effect in momentum space. Furthermore, our numerical simulations (not shown) reveal that reducing the driving strength or increasing the driving frequency can also boost the self-trapping effect, thereby resulting in a persistent and stable current.

\section{The chaos signature of phase transition}
In Fig. \ref{q3} we have calculated the deviation between two temporally evolving states from two slightly different initial conditions according to the full GP equation, showing that the deviation reaches a maximum at the phase transition point. To further confirm that the chaotic behaviour occurs in the vicinity of the phase transition point, we can quantitatively describe the chaotic signature by the simple three-mode model (TMM) and study the sensitivity of the system dynamics to the initial condition. To do this, a set of small quantities $\boldsymbol{\epsilon}=(\delta A,\delta B,\delta C)$ is defined to describe the  adjacent trajectories of two nearby points moving forward in phase space.  Linearizing Eq. (\ref{eq8}) we get the evolution equations of $\boldsymbol{\epsilon}$,
\begin{align}\label{eq9}
	i\delta \dot{A}=&\frac{g}{2\pi}\big [2|A|^2\delta A+A^2\delta A^{*}+B^2\delta C^{*}+2BC^{*}\delta B \nonumber\\
	&+2(|B|^2\delta A+AB^{*}\delta B+AB\delta B^{*})+2(|C|^2\delta A 			         \nonumber \\&+AC^{*}\delta C+AC\delta C^{*})\big ]
	+\frac{\delta A}{2}+\frac{ik\sin(\omega t)}{2}\delta B, \nonumber \\
	i\delta \dot{B}=&\frac{g}{2\pi}\big[2|B|^2\delta B+B^2\delta B^{*}+2(AB^{*}\delta C+AC\delta B^{*} \nonumber \\
	&+B^{*}C\delta A)+2(AB\delta A^{*}+A^{*}B\delta A+|A|^2\delta B)                           	    \nonumber \\&+2(CB\delta C^{*}+C^{*}B\delta C+|C|^2\delta B)\big ]
	\nonumber \\&-\frac{iK\sin(\omega t)}{2}(\delta A-\delta C), \nonumber \\
	i\delta \dot{C}=&\frac{g}{2\pi}\big [2|C|^2\delta C+C^2\delta C^{*}+B^2\delta A^{*}+2A^{*}B\delta B \nonumber\\
	&+2(|A|^2\delta C+AC\delta A^{*}+A^{*}C\delta A)+2(|B|^2\delta C\nonumber           			   \\&+BC\delta B^{*}+B^{*}C\delta B)\big ]
	+\frac{\delta C}{2}-\frac{ik\sin(\omega t)}{2}\delta B.
\end{align}

We use  $\delta P_{-1}(t)=|A+\delta A|^2-|A|^2$ to characterize the deviation of population distribution. Then its logarithm $\ln(\delta P_{-1} )$ shows the tendency of the perturbation (an increasing tendency generally implies the existence of chaos) and the logarithmic slope of the perturbation $\delta P_{-1} $ with respect to time is defined as the Lyapunov exponent, which is explicitly given by
\begin{equation}\label{eq10}
	\uplambda_{LE}=\lim_{t\to \infty}\frac{1}{t}\ln\frac{\delta P_{-1}(t)}{\delta P_{-1}(0)}.
\end{equation}
\begin{figure}[htp]
	\center
	\includegraphics[width=7cm]{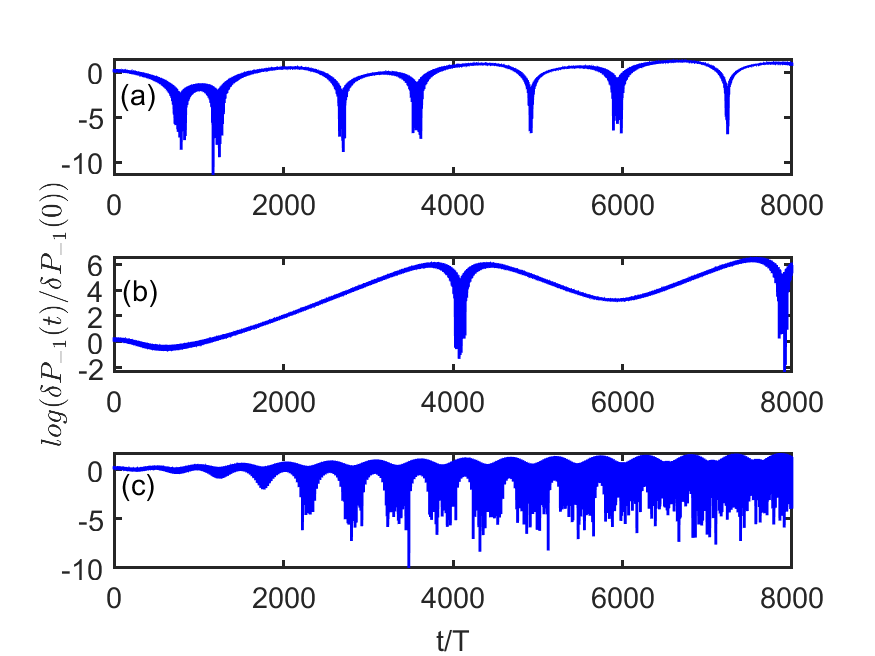}
	\caption{Trajectory curves of $\ln \big(\delta P_{-1}(t)/\delta P_{-1}(0)\big )$ versus time. (a):  below the phase transition point, $g=0.05$, there is no rising trend for the curve. (b): at the phase transition point,  $g=0.089$, the  $\ln \big(\delta P_{-1}(t)/\delta P_{-1}(0)\big )$ curve rises from zero to nearly 6, indicating the exponential divergence of nearby trajectories. (c): above the phase transition point, $g=0.2$, there is no rising trend either. The other parameters are $K=2, \omega=10$.} \label{q5}
\end{figure}

\begin{figure}[htp]
	\center
	\includegraphics[width=7cm]{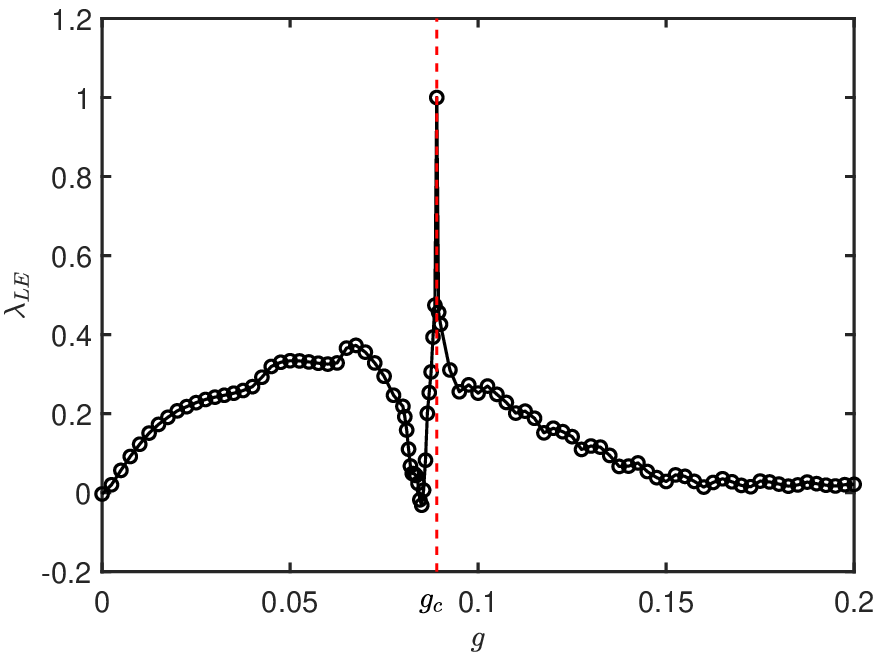}
	\caption{Lyapunov exponent $\uplambda_{LE}$ versus nonlinearity strength $g$ within the parameter range between 0 and 0.2. The other parameters are $K=2,\omega=10$.} \label{q6}
\end{figure}

\begin{figure}[htp]
	\center
	\includegraphics[width=7cm]{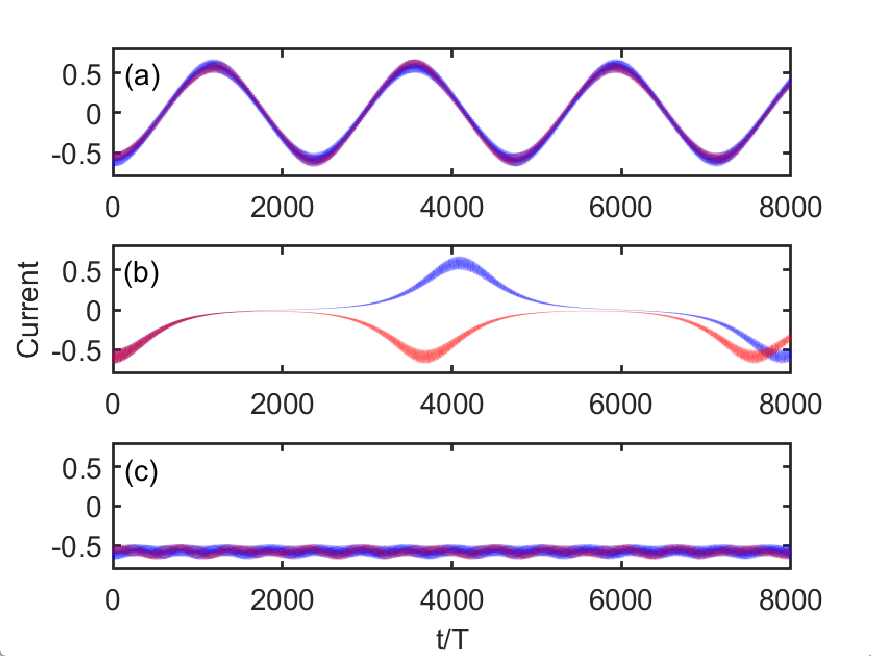}
	\caption{Time evolution of the current predicted by the TMM, starting from the initial state (blue line) $A(0)=B(0)=\frac {1}{\sqrt{2}}$, $C(0)=0$ and its slightly perturbed state (red line) by $\boldsymbol{\epsilon}=(10^{-3},10^{-3},10^{-3})$. (a): $g=0.05$ below the phase transition point. (b): $g=0.089$ at the phase transition point. (c): $g=0.2$ above the phase transition point. The other parameters are $K=2$ and $\omega=10$.}\label{q7}
\end{figure}

Selecting initial values $\boldsymbol{\epsilon}=(10^{-20},10^{-20},10^{-20})$, we numerically integrate Eq. (\ref{eq9}) using a  Runge-Kutta method, parallel to the numerical integration of Eq. (\ref{eq8}) with initial condition $A(0)=B(0)=\frac {1}{\sqrt{2}},C(0)=0$. The plot of $\ln(\delta P_{-1}(t)/\delta P_{-1}(0))$ versus time $t$ is shown in Fig. \ref{q5}(a)-\ref{q5}(c) with three different values of $g$. When the nonlinearity strength $g$ is below or above the phase transition point, the curves of  $\ln(\delta P_{-1}(t)/\delta P_{-1}(0))$ in Fig. \ref{q5}(a) and \ref{q5}(c) show no rising trend, whereas at the phase transition point, the curve of  $\ln(\delta P_{-1}(t)/\delta P_{-1}(0))$  in Fig. \ref{q5}(b) shows an upward trend from 0 to nearly 6, implying the exponential divergence of nearby trajectories. In addition, we have fitted the Lyapunov exponent versus $g$ for parameter ranges of 0 and 0.2 in Fig. \ref{q6}. Since the actual Lyapunov exponent is very small, in order to guide the reader's eye,  in Fig. \ref{q6} we use the ratio of the true Lyapunov exponent of any parameter to the Lyapunov exponent of the phase transition point.  Fig. \ref{q6} shows that the Lyapunov exponent reaches its maximum at the transition point, confirming that chaos can be used as an indicator of the current quantum phase transition. The sensitivity of the current dynamics to the initial conditions is discussed more intuitively in Fig. \ref{q7}, where at the phase transition point we observe the striking deviation of the currents starting from the initial state $A(0)=B(0)=\frac {1}{\sqrt{2}}$, $C(0)=0$ and its perturbed state by $\boldsymbol{\epsilon}=(10^{-3},10^{-3},10^{-3})$, whereas on both sides of the transition point there is a coincidence between the currents.
\\

\begin{figure}[htp]
	\center
	\includegraphics[width=7cm]{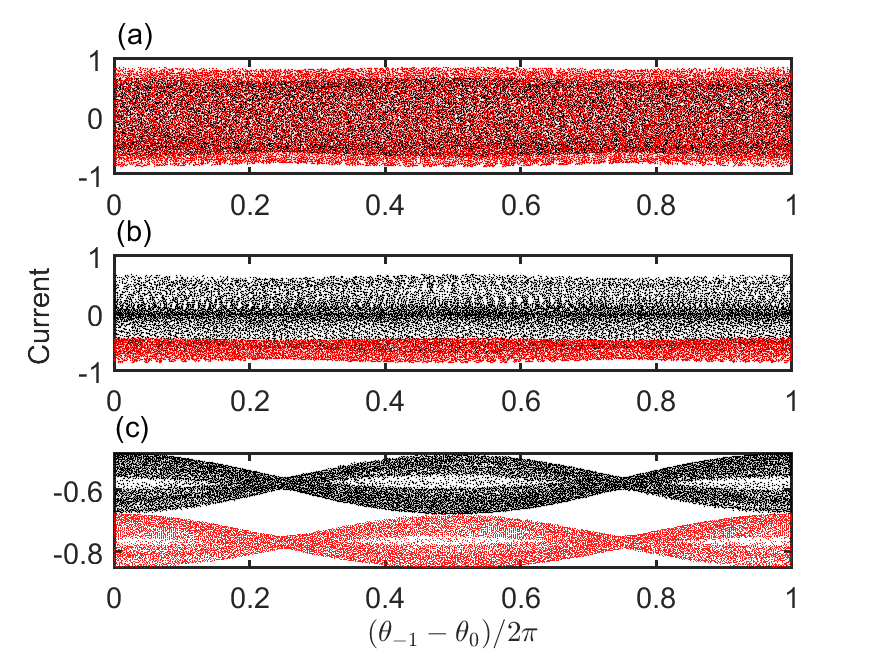}
	\caption{Phase-plane portrait of the two dynamical variables,  the current and the phase difference between the modes $\ket{-1}$ and $\ket{0}$, with two initial conditions. The black dots are for the EWS initial condition, $A(0)=B(0)=\frac {1}{\sqrt{2}}$, $C(0)=0$, and the red dots are for the UEWS initial condition, $A(0)=\sqrt{0.7}$, $B(0)=\sqrt{0.3}$, $C(0)=0$. (a): $g=0.05$ below the phase transition point. (b): $g=0.089$ at the phase transition point. (c): $g=0.2$ above the phase transition point. The other parameters are $K=2$ and $\omega=10$.} \label{q8}
\end{figure}

Another advantage of the TMM is that the current dynamical behaviour can be quite conveniently represented in terms of a phase portrait by choosing different initial conditions. In Fig. \ref{q8} we show the phase portrait of the two dynamical variables, i.e. the current and the phase difference between the modes $\ket{-1}$ and $\ket{0}$, with two asymmetric initial states: one is the equally weighted superposition (EWS) between the modes $\ket{-1}$ and $\ket{0}$ and the other is the unequally weighted superposition (UEWS) between the modes $\ket{-1}$ and $\ket{0}$, specifically $A(0)=\sqrt{0.7}$, $B(0)=\sqrt{0.3}$, $C(0)=0$. As shown in Fig. \ref{q8}(a), when the nonlinearity strength is below the phase transition point, the phase diagram is symmetric, suggesting no directed transport for both initial states. At the transition point, the trajectory for the initial state of the EWS (the back line) remains symmetric, while the trajectory for the initial state of the UEWS (the red line) loses its symmetry, and it is constrained to be asymmetric in the vicinity of the line $I(t)=-0.6$, as illustrated in Fig. \ref{q8} (b). As the nonlinear parameter crosses over the transition point,  the interesting thing observed in Fig. \ref{q8}(c) is that the initial current will be well maintained regardless of the choice of initial states.
\\

\begin{figure}[htp]
	\center
	\includegraphics[width=7cm]{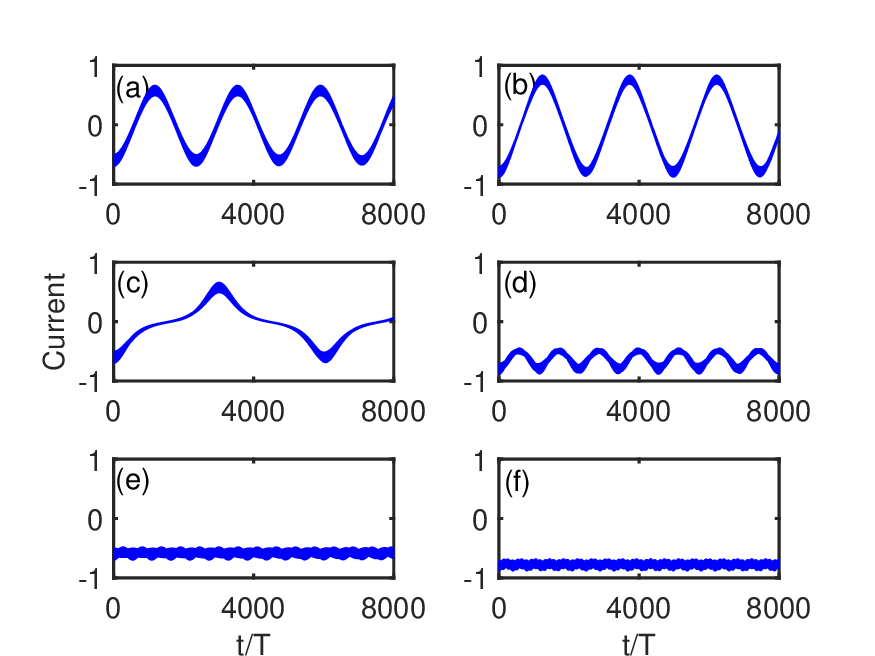}
	\caption{Time evolution of the current obtained from the simulation of the full GP equation (\ref{eq1}), starting from the EWS initial condition (left column) and the UEWS initial condition (right column). Top panels: $g=0.05$ below the phase transition point; Middle panels: $g=0.089$ at the phase transition point; Bottom panels: $g=0.2$ above the phase transition point. The other parameters are $K=2,\omega=10$.} \label{q9}
\end{figure}

Finally, we calculate the currents $I(t)$ by going back to the GP equation (\ref{eq1}) with the same two initial states as in Fig. \ref{q8}. As shown in Fig. \ref{q9}, the numerical results of Eq. (\ref{eq1}) are consistent with those predicted by the three-mode model. Direct numerical simulation of the GP equation confirms that (i) below the phase transition, the TAC is zero regardless of the initial state; (ii) at the phase transition point, whether or not there is current generation depends on the initial conditions. The higher the momentum expectation value in the initial state, the more likely it is that directed transport will occur; (iii) beyond the transition point, the asymmetry of the momentum distribution in the initial state is well preserved during the dynamical evolution, and thus the initial current is correspondingly maintained. This is due to the self-trapping effect induced by the nonlinearity.
\section{Conclusion}
To conclude, we present a comprehensive analysis of current generation in a nonlinear periodically driven system involving a Bose-Einstein condensate on a ring. We show that the sharp transition between the vanishing current regime and the nonvanishing current regime can be achieved by adjusting the nonlinearity or the driving amplitude and frequency. We demonstrate that the directed current presented here is not generated by the broken spatio-temporal symmetries of the system, but rather by the self-trapping effect in the momentum space. By taking advantage of the self-trapping effect, we can achieve a novel ratchet-like transport that allows the condensate to move at an almost constant speed by reducing the driving strength or increasing the driving frequency, even when the nonlinearity strength is relatively small.  In particular, we find that the Lyapunov exponent reaches a maximum at the current phase transition point, demonstrating the chaos manifestation of the current transition. These results provide a unique way to control of the directed current, in particular to advance the understanding of fundamental physics of the quantum ratchet effect induced by self-trapping in the nonlinear system.

\acknowledgments
The work was supported by  the National
Natural Science Foundation of China (Grant Nos. 12375022, 11975110), the Natural Science Foundation of Zhejiang Province (Grant No. LY21A050002), Zhejiang Sci-Tech University Scientific Research
Start-up Fund (Grant No. 20062318-Y), and  Jiangxi Provincial Natural Science Foundation (Grant No. 20232BAB201008).

\end{document}